\documentclass[aps,twocolumn,showpacs,superscriptaddress,citeautoscript,reprint,prb,floatfix]{revtex4-1}

\usepackage[utf8]{inputenc}
\usepackage{amsmath}
\usepackage{mathtools}
\usepackage{amsfonts}
\usepackage{amssymb}
\usepackage{graphicx}
\usepackage{hyperref}
\usepackage{bbold}
\usepackage{siunitx}
\usepackage{color}

\newcommand{\revision}[1]{\textcolor{black}{{#1}}}

\renewcommand{\vec}[1] {{\mathbf{#1}}}
\newcommand{\la}{\lambda}
\newcommand{\ham}{\mathcal{H}}
\newcommand{\ce}[1] {$\mathrm{#1}$}
\newcommand{\abs}[1] {{\lvert}#1{\rvert}}
\newcommand{\bra}[1] {\langle #1{\rvert}}
\newcommand{\ket}[1] {\lvert #1{\rangle}}

\begin{document}

\title{Electric field manipulation of surface states in topological semimetals}

\author{Yuriko Baba}
\email{yuribaba@ucm.es}
\affiliation{Instituto de Estructura de la Materia, IEM-CSIC, Serrano 123, E-28006 Madrid, Spain}
\affiliation{GISC, Departamento de F\'{\i}sica de Materiales, Universidad Complutense, E-28040 Madrid, Spain}

\author{\'Alvaro D\'{\i}az-Fern\'andez}
\affiliation{GISC, Departamento de F\'{\i}sica de Materiales, Universidad Complutense, E-28040 Madrid, Spain}

\author{Elena D\'{\i}az}
\affiliation{GISC, Departamento de F\'{\i}sica de Materiales, Universidad Complutense, E-28040 Madrid, Spain}

\author{Francisco Dom\'{\i}nguez-Adame}
\affiliation{GISC, Departamento de F\'{\i}sica de Materiales, Universidad Complutense, E-28040 Madrid, Spain}

\author{Rafael A. Molina}
\email{rafael.molina@csic.es}
\affiliation{Instituto de Estructura de la Materia, IEM-CSIC, Serrano 123, E-28006 Madrid, Spain}

\begin{abstract}

We investigate the consequences of applying electric fields perpendicularly to thin films of topological semimetals. In particular, we consider Weyl and Dirac semimetals in a configuration such that their surface Fermi arcs lie on opposite edges of the films. 
We develop an analytical approach based on perturbation theory and a single-surface approximation and we compare our analytical results with numerical calculations. The effect of the electric field on the dispersion is twofold: it shifts the dispersion relation and renormalizes the Fermi velocity, which would, in turn, have direct effects on quantum transport measurements. Additionally, it modifies the spatial decay properties of surface states which will impact the connection of the Fermi arcs in opposite sides of a narrow thin film.
\end{abstract}

\maketitle

\section{Introduction}
\label{sec:intro}

Topological materials have attracted great interest in the last decade since they exhibit new fundamental phenomena and hold great promise for far-reaching technological applications. A hallmark of topological materials is quantized response functions and the existence of protected gapless surface states, which arise due to the nontrivial topology of the bulk states by virtue of the bulk-boundary correspondence~\cite{Shen2017_Book}. Nontrivial topology can be characterised by topological invariants according to the symmetry class of the system enabling a complete classification in both gapped and gapless systems~\cite{Chiu2016}. The gapped case was the first under study, starting the fruitful field of topological insulators (TIs)~\cite{Bernevig2013_Book,Shen2017_Book}. On the other hand, gapless systems assemble the family of topological semimetals (TSMs), where the valence and conduction bands only touch at a zero-measure set of points in the Brillouin zone (BZ). In particular, topological Weyl and Dirac semimetals (WSMs and DSMs) are three-dimensional phases of matter in which these isolated touching points, dubbed Weyl and Dirac points respectively, are protected by topology and symmetry~\cite{Burkov2018,Armitage2018}.

In WSMs, near the Weyl node, the dispersion relation appears as a three-dimensional analogous to graphene and can be described by an anisotropic version of the Weyl equation. The low-energy quasiparticles behave, then, as relativistic Weyl fermions. The conduction and valence bands are individually non-degenerate and hence either time reversal symmetry or inversion symmetry need to be broken. The Weyl nodes are monopoles of Berry curvature and the charge associated with them is called chirality \cite{Burkov2018, Armitage2018}. Due to the Nielsen-Ninomiya no-go theorem \cite{Nielsen1983}, Weyl nodes always come in pairs and can annihilate only in pairs. Hence, the robustness of the Weyl nodes is quantified by the separation of the nodes in reciprocal space. For a given momentum between the Weyl nodes, the 3D WSM can be mapped onto a 2D TI \cite{Shen2017_Book, Turner2013}. This correspondence leads to non trivial Chern numbers and to protected states localized at the surface. The protected states, called Fermi arcs, lie in open contours at the Fermi energy. Several materials have been predicted and confirmed to be topological WSMs with Fermi arcs by means of ARPES experiments. The most representative is the \ce{TX} family where \ce{T= Ta/Nb} and \ce{X= As/P}, \ce{TaAs} being a particular case \cite{Huang2015, Weng2015, Xu2015, Lv2015a}. These materials belong to the so-called Type I WSMs and are characterised by a discrete point-like Fermi surface. However, it is possible to have anisotropies in the dispersion so that the Fermi surface is not a point but it becomes an open surface, thereby leading to electron and hole pockets. Materials within this category are referred to as Type II Weyl semimetals~\cite{Huang2016}.

The DSMs are obtained when both time reversal and inversion symmetry are present. Consequently, the Dirac points have a four-fold degeneracy and the net Chern number of the nodes is zero. Each Dirac point can be constructed by imposing two Weyl nodes with opposite chirality and, in order to be topologically protected, it must be stabilised by additional symmetries as the up-down parity symmetry \cite{Gorbar2015R, Gorbar2015} or by space-group symmetries \cite{Young2012, Gibson2015, Steinberg2014}. This is the case of compounds like \ce{A_3 Bi} where \ce{A = Na, K, Rb} and \ce{Cd_3As_2}, in which ARPES experiments ratify the existence of the Fermi arcs \cite{Neupane2014, Liu2014a, Zhang2014, Xu2015}. The manufacturing of high quality thin films \cite{Hellerstedt2016,Zhang2014} and ultrathin films of \ce{Na_3 Bi} \cite{Collins2018} has made this material one of the most promising candidates for technological applications of topological properties. For example, monolayer and bilayer films of \ce{Na_3 Bi} have bulk bandgaps greater than \SI{300}{\milli\electronvolt},  suggesting that topological properties in these thin films will survive at room temperature~\cite{Niu2017,Collins2018}.

The topological response of WSMs and DSMs comprises the manifestation of the chiral anomaly in a large negative magnetoresistance, in the presence of both electric and magnetic fields, and in an anomalous Hall effect (in WSMs), due to the transport of the surface states.~ \cite{Shen2017_Book, Armitage2018} Therefore, the renormalization of the Hamiltonian parameters due to time-dependent external fields has been the subject of intense research.~\cite{Narayan2015,Chan2016} In this case, new surface states with interesting properties may appear.~\cite{Gonzalez2016} Overall, the understanding of the effect of external fields on the topological phases and transport phenomena is a field of great interest both from a first principles standpoint and from the perspective of possible applications. Hence, a special interest resides in understanding the effect of the external fields in the most direct manifestation of topology, i.e. the surface states. 

The present manuscript studies the effects of an electric field applied perpendicularly to the surface and to the direction joining the line of nodes. In Sec. \ref{sec:Models}, we introduce the theoretical framework: a minimal model for a generic WSM and a low-energy effective model for the DSM \ce{Na_3Bi}. Section \ref{sec:fField} describes the effect of an external electric field applied perpendicularly to the direction along which the two nodes are aligned. We provide analytic results obtained by means of perturbation-theory techniques. In order to extend our results to non-perturbative regimes, we provide numerical calculations that match the analytic results.  Finally, in Sec. \ref{sec:conc} we finish with some conclusions and a brief analysis of the experimental feasibility for measuring the predicted behavior in thin films of the topological semimetal \ce{Na_3Bi}.


\section{Model Hamiltonians} \label{sec:Models}

\subsection{Minimal model for a WSM and DSM} \label{subsec:WeylMM}

The minimal setting for describing a Weyl semimetal consists of  two Weyl points that are realised in a time-reversal-symmetry-breaking scenario while preserving inversion symmetry \cite{Armitage2018, Shen2017_Book}. A generic low-energy Hamiltonian that meets these requirements can be written as~\cite{Gonzalez2017}
\begin{equation} \label{eq:ModMin:Ham}
    \mathcal{H}_{\zeta}= (m_0-m_1 \vec{k}^2)\sigma_z +vk_z\sigma_x+ \zeta v k_y\sigma_y \ ,
\end{equation}
where $\zeta$ is $\pm 1$ depending on the chirality,  $\sigma_i$ with $i=x,y$ and $z$ are the Pauli matrices and $\vec{k}=(k_x,k_y,k_z)$ are momentum operators. In addition, $m_i$ with $i = \{ 0,1\}$ account for mass parameters and $v$ is the Fermi velocity. 
From the previous Hamiltonian, a DSM can be build from two copies with opposite chirality, which are time-reversal partners. Thus, if no chirality-mixing term is considered, the Hamiltonian can be written in the following block diagonal form
\begin{equation} \label{eq:HamD}
    \ham_{D}(\vec{k})=\left( 
    \begin{array}{cc}
    \ham_{\zeta=+1}(\vec{k}) & 0 \\
    0 & \ham_{\zeta=-1}(\vec{k})
    \end{array}
    \right) ~.
\end{equation}

In bulk, the dispersion relation of  Hamiltonian~(\ref{eq:ModMin:Ham}) is given by
\begin{equation}  \label{eq:ModMin:bEns}
    E_b = \pm \sqrt{\left(m_0 - m_1 \abs{\vec{k}}^2\right)^2 + v^2 (k_y^2 + k_z^2)} \ .
\end{equation}
The valence band and the conduction band touch at the aforementioned Weyl points, located at $\vec{k_W}_\pm= (\pm \sqrt{m_0/m_1},0,0)$. The Weyl points are monopoles of Berry curvature and have Chern number equivalent to their chirality. The nonzero Chern number leads, according to the bulk-boundary correspondence, to surface states named Fermi arcs \cite{Hatsugai1993}.

For the analytic approach, we consider semi-infinite geometries in the perpendicular direction to the node separation by introducing a single surface in the $Z$-direction. In order to explore how the location of the boundary affects the dispersion, we shall let the boundary sit at $z=-\eta w$, where $\eta=\pm 1$ indicates the position of the surface with respect to the plane $z=0$ and $w>0$. Bear in mind that we are considering a single surface but are allowing for both signs of $\eta$ to see its manifestation in the dispersion. Surface states are obtained from Hamiltonian (\ref{eq:ModMin:Ham}) by using the \textit{ansatz} $\psi_{s} \sim e^{i \vec{k}_{\bot}\cdot\vec{r}} e^{-\lambda (\eta z+ w)} \Phi$, where $\Phi$ is a constant spinor, $\vec{k}_{\bot}=(k_x,k_y,0)$ and $\lambda$ is a complex number with a nonzero real part. For simplicity, from now on we make implicit the plane-wave dependence $ \exp(i \vec{k}_{\bot}\cdot\vec{r})$. A solution satisfying Dirichlet boundary conditions is given by~\cite{Gonzalez2017}
\begin{equation} \label{eq:ModMin:SurfS}
\psi_ {s} = \frac{A_s}{\sqrt{2}} \left(e^{-\la_1 ( \eta z + w)}-  e^{-\la_2 (\eta z + w)} \right) 
\begin{pmatrix}
 1\\ \eta i
\end{pmatrix} \ , 
\end{equation} 
where $A_s$ is a normalization factor and
\begin{equation}
    \la_{1}= \Delta + \sqrt{F} \ ,
    \qquad
    \la_{2}= \Delta - \sqrt{F} \ .
    \label{eq:ModMin:lam}
\end{equation}
Here we have defined  
\begin{equation}
    \Delta\equiv v/(2m_1) \ ,\qquad F \equiv k_x^2+k_y^2-R^2+\Delta ^2 \ ,
    \label{eq:ModMin:F}
\end{equation}
with $R \equiv \sqrt{m_0/m_1}$. Surface states occur whenever $\Re[\lambda_{1,2}]>0$, which implies $F<\Delta^2$. Equivalently, surface states are restricted to a circle of radius $R$ in momentum space
\begin{equation} \label{eq:ModMin:CondS}
    {k_x^2+ k_y^2}< R^2\ .
\end{equation}

At a given $k_x$ that supports surface states, the dispersion relation is linear in $k_y$ and depends on $\zeta$ and on the position of the surface
\begin{equation} \label{eq:ModMin:SurfE}
    E_s = \eta \zeta v k_y \ .
\end{equation}

\begin{figure}
   \centering
    \includegraphics[width=0.35\textwidth]{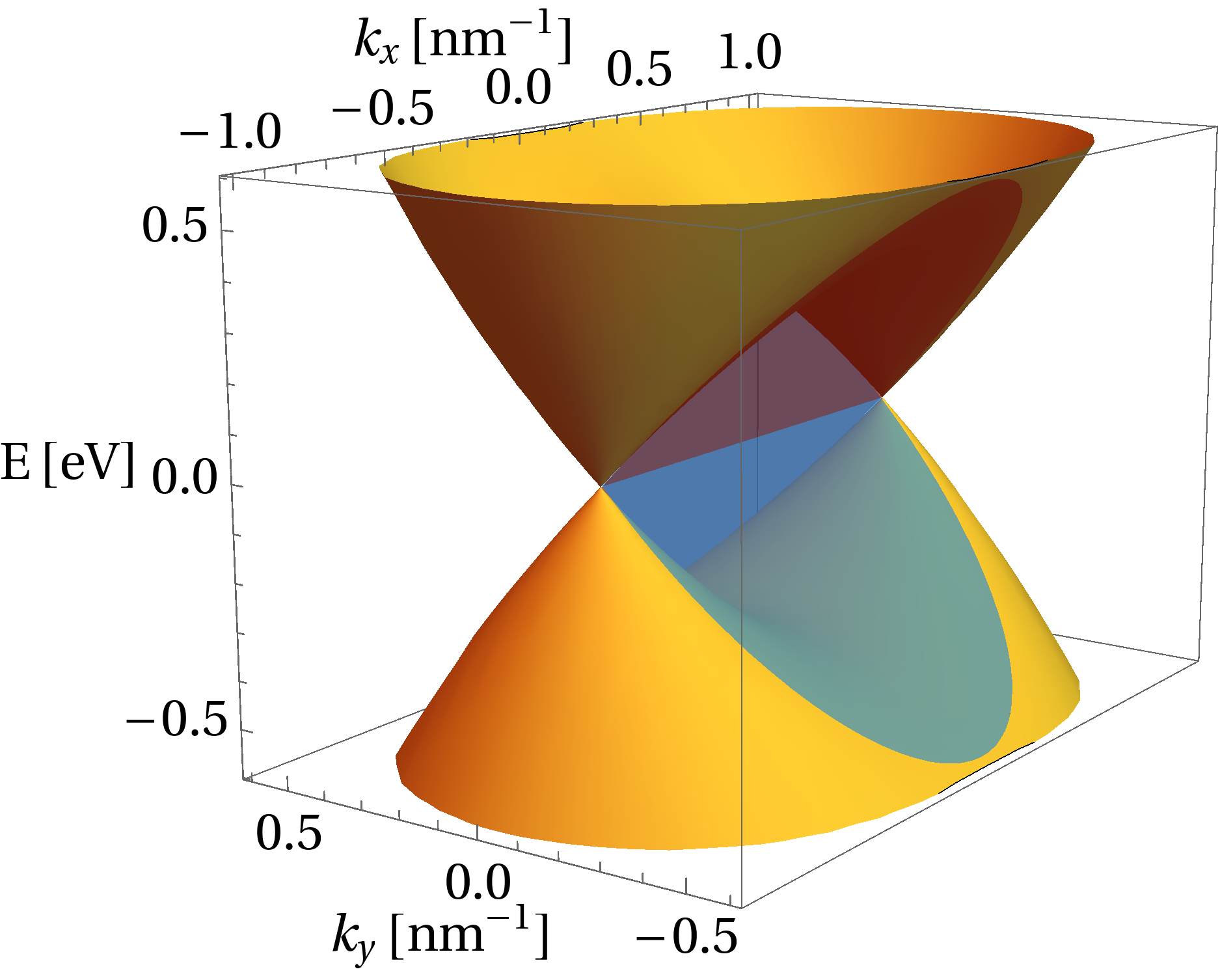}
    \caption{Dispersion of a system with a boundary at $z=-\eta w$. Bulk bands are depicted in orange, whereas surface bands are shown in opaque red and light-blue for $\eta \zeta = -1$ and $\eta \zeta = 1$, respectively. The parameters chosen for the plot are $m_0=\SI{0.35}{\electronvolt}$, $m_1=\SI{1.0}{\electronvolt\nano\meter\squared}$ and
    $v=\SI{1.0}{\electronvolt \nano\meter}$.}
\label{fig:ModMin:Disp}
\end{figure}

This dispersion is represented in Fig. \ref{fig:ModMin:Disp} together with the bulk bands. From the dispersion relation we can give another interpretation to the condition of existence of surface states. Indeed, the region \eqref{eq:ModMin:CondS} defines a circle in the plane $(k_x,k_y)$ outside which the surface states bands become tangent to the bulk states dispersion. This leads to a hybridisation between surface states and bulk states which prevents us to use the ansatz and to have Fermi arcs.

It is interesting to point out that there are two types of surface states, depending on whether $\lambda_{1,2}$ have or not an imaginary part. If they do, the exponential decay will be accompanied by oscillations. From equation~(\ref{eq:ModMin:F}) we can define a transition radius as
\begin{equation}
    r_{\mathrm{trans}}^2 \equiv R^2-\Delta^2 \ .
    \label{eq:ModMin:rtrans}
\end{equation}
Then, we can see that oscillatory states live in a circle of radius $r_{\mathrm{trans}}$, whereas purely exponential states are found in a planar ring of inner radius $r_{\mathrm{trans}}$ and outer radius $R$. Hereafter we shall denote the oscillatory states as type A and purely exponential states as type B. Since the type B states are closer to the bulk states in energy, these have longer decay lengths than type A~\cite{Benito-Matias2019, Gonzalez2017}. Notice that $r_{\mathrm{trans}}^2$ can either be positive or negative, depending on the parameters of the model. If positive, both type A and B states exist. However, if negative, only type B arise. In the following, we will loosely dub Weyl semimetals hosting type A states as type A Weyl semimetals, carefully remembering that these semimetals also host type B states.

\subsection{Model for the Dirac semimetal \ce{Na_3Bi}} \label{subsec:Na3Bi}

In order to elucidate the generality of our results and to have a more realistic approach to the problem, we will also study the case of a low-energy effective Hamiltonian that describes \ce{A_3Bi} (\ce{A= Na, K, Rb}) \cite{Wang2012} and \ce{Cd_3As_2} \cite{Wang2013} around the $\Gamma$ point. These compounds have a single band inversion occurring near the $\Gamma$ point that has been observed by ARPES measurements \cite{Neupane2014, Liu2014a, Zhang2014, Xu2015}. For concreteness, we restrict our analysis to the case of \ce{Na_3Bi}. After a density functional theory analysis, the Hamiltonian can be cast in the form of a DSM equivalent to Eq.~\eqref{eq:HamD} in which the corresponding WSM Hamiltonian is replaced by~\cite{Wang2012}
\begin{equation} \label{eq:Na3Bi:HamW}
    \mathcal{H}_{\zeta} = \epsilon_0(\vec{k}) \mathbb{1}_{2} + M(\vec{k}) \sigma_z +  v (\zeta k_x \sigma_x- k_y\sigma_y) \ ,
\end{equation}
where $\epsilon_0(\vec{k})=c_0+c_1 k_z^2+c_2(k_x^2+k_y^2)$ and $ M(\vec{k})=m_0-m_1 k_z^2-m_2(k_x^2+k_y^2)$, $c_i$ and $m_i$ with $i = \{ 0, 1, 2\}$ being constant parameters and mass terms respectively. Notice that, except for the diagonal contribution, $\ham_{\zeta}$ is a rotated version of~(\ref{eq:ModMin:Ham}) that allows for anisotropy along the $Z$-direction. Threefold rotational symmetry implies that chirality-mixing terms are of order $\mathcal{O}(k^3)$ and we will neglect them in our analysis, thereby effectively decoupling the two Dirac nodes~\cite{Wang2012}. This type of Dirac semimetal is also referred to as a $\mathbb{Z}_2$ WSM \cite{Gorbar2015R, Gorbar2015}.
 
The surface states of the model have been worked out in detail in Ref.~\onlinecite{Benito-Matias2019}. By placing a surface termination at $y=-\eta w$ and imposing Dirichlet boundary conditions as before, the surface states take the form
\begin{subequations}
\begin{align}\label{eq:Na3Bi:SurfS}
    \psi_{s} &= A_s (e^{-\la_1 (\eta y+w)}- e^{-\la_2 (\eta y+w)}) \Phi_{\eta}\ , \\
    \Phi_{\eta} &=
    \left( \begin{array}{c}
        \eta  \\ \sqrt{\frac{m_2-c_2}{m_2+c_2}}
        \end{array} \right)\ ,
\end{align} 
\end{subequations}
where $A_s$ is a normalisation factor. Here, $\lambda_{1,2}$ are defined in equation~(\ref{eq:ModMin:lam}), where $\Delta$ and $F$ are now given by
\begin{subequations}
\begin{align}
F & \equiv (k_x + \eta \zeta k_{x,0})^2 +\frac{k_z^2}{(m_2/m_1)} + \Delta^2- R^2\ , \\
\Delta & \equiv \frac{v}{2\sqrt{m_2^2-c_2^2}} \ ,
\end{align}
\end{subequations}
being $k_{x,0} \equiv {c_2}\Delta/{m_2}$ and $R^2 \equiv {m_0}/{m_2}+ \Delta^2\left({c_2}/{m_2}\right)^2$. In contrast to the previous case, the diagonal term $\epsilon_0(\vec{k})\mathbb{1}_2$ leads to a dispersion that is no longer flat along the $Z$-direction (recall that this model is rotated with respect to the minimal model presented above). Instead, surface states now have the following dispersion
\begin{equation} \label{eq:Na3Bi:SurfE}
E_s = \varepsilon(k_z)
+\eta \zeta v C_3k_x \ ,
\end{equation}
where
\begin{equation}
    \varepsilon(k_z) = C_1+C_2 k_z^2 \ .
    \label{eq:Na3Bi:dispz}
\end{equation}
In these two equations, $C_1,C_2$ and $C_3$ are a combination of the Hamiltonian parameters and are given by $C_1 = c_0+c_2 {m_0}/{m_2}$,  $C_2=c_1-c_2 {m_1}/{m_2}$ and $C_3 = \sqrt{1-c_2^2/m_2^2}$. Notice that if we set all $c_i=0$ and $m_1=m_2$, all expressions reduce to those obtained in the previous section and we recover a flat band behaviour along the $Z$-direction. The classification in type A and type B states remains the same, that is, type B states are purely exponential whereas type A have an oscillatory component to the exponential decay. In \ce{Na_3Bi} only type B states arise as can be demonstrated by introducing into the above definitions the parameters listed in table~\ref{tab:Na3Bi}. 
\begin{table}[h]
    \begin{tabular}{|l|l|l|} 
    \hline
         $c_0 = \SI{-0.06382}{\electronvolt^{2}}$  &
         $m_0 = \SI{-0.08686}{\electronvolt^{2}}$ &
         \\
         $c_1 = \SI{8.7536}{\electronvolt \angstrom^2}$ &
         $m_1 = \SI{-10.6424}{\electronvolt \angstrom^2}$ &
         $v = \SI{2.4598}{\electronvolt \angstrom}$
         \\
         $c_2 = \SI{ -8.4008}{\electronvolt \angstrom}$ &
         $m_2 = \SI{-10.3610}{\electronvolt \angstrom}$ &
         \\ \hline 
    \end{tabular}

    \caption{Parameters for the Hamiltonian of \ce{Na_3 Bi} extracted from Ref.~\onlinecite{Wang2012}.}
    \label{tab:Na3Bi}
\end{table}


\section{Electric field} \label{sec:fField}

In this section we study the effect of an external electric field applied perpendicularly to the surface. The surface configurations correspond to those studied in the previous section. Earlier works in Dirac materials have proven that a renormalization of the Fermi velocity of the surface states occurs in the presence of a perpendicular electric field \cite{Diaz-Fernandez2017TuningField, Diaz-Fernandez2017Quantum-confinedJunctions}. Therefore, we expect similar effects in WSMs and DSMs. We approach the problem in two ways: i) by way of perturbation theory (PT) to obtain analytic results and ii) by performing numerical calculations based on the Python package Kwant~\cite{Groth2014Kwant:Transport}. In this way, we can compare the validity of the analytic results and study the non-perturbative regime.

\subsection{Minimal model for a WSM and DSM} \label{subsec:fMM}

We begin by considering the minimal model under the influence of an electric field applied along the $Z$-direction, perpendicular to the surface. 
Thus, its perturbation reads
\begin{equation}
    \mathcal{H}_f= efz\mathbb{1}_2 \ ,
\end{equation}
where $e$ is the elementary electric charge and $f$ is the external electric field. The first order correction in perturbation theory is given by
\begin{equation}
\label{eq:fMM:Pert1st}
    \delta E^{1}_s = \bra{\psi_s^0}\ham_f \ket{\psi_s^0} \ ,
\end{equation} 
where $\psi_s^0$ are the surface states in the absence of an electric field as defined in Eq.~\eqref{eq:ModMin:SurfS}. From Eq.~\eqref{eq:fMM:Pert1st}, the correction to the energy is given by
\begin{equation} \label{eq:fMM:E1}
    \delta E^{1}_s = \eta e f  \left[ \Gamma_0 + \Gamma_1(k_x, k_y) \right]\ ,
\end{equation}
where we have defined
\begin{subequations}
\begin{align}
\Gamma_0 & \equiv \frac{v}{2 m_0}+\frac{m_1}{v}-w  \label{eq:fMM:Gamma0} \ ,\\
\Gamma_1 (k_x, k_y) &  \equiv \frac{v \left(k_x^2+k_y^2\right)}{2 m_0 \left(R^2- k_x^2-k_y^2\right)} \label{eq:fMM:Gamma1} \ .
\end{align}
\end{subequations}
Notice that these corrections are independent of chirality and of the surface state type. Equation~(\ref{eq:fMM:E1}) reveals that there is a constant shift in energy, together with a momentum dependent term due to the perturbation. The reasoning behind this is that the decay lengths are momentum dependent as well. In fact, we can try to provide with a simple argument. Let us set zero potential at $z=0$ such that it is $-ef\eta w$ at the surface termination. Then, one may consider a surface far from $z=0$, so that $\Gamma_1$ becomes negligible with respect to $\Gamma_0$, so long as we consider low momenta. Moreover, all terms in $\Gamma_0$ are negligible except for $-w$ if the surface is sufficiently far from $z=0$. In that case, the correction is simply $-ef\eta w$. This means that the potential is locally acting on the surface states by simply lowering their energy by an amount equal to the value of the potential at the surface. More generally, in first order perturbation theory we are calculating $ef\langle z\rangle$, which essentially amounts to calculating the expectation value of the position in the unperturbed surface state. Hence, both $\Gamma_1$ and the terms in $\Gamma_0$ that accompany $w$ represent a correction to $-\eta w$ due to the fact that the surface state has some extension and penetrates slightly into the bulk. Notice that $\Gamma_1$ presents a first order pole at the momenta located in the circle of radius $R$. This is consistent with the fact that upon approaching the edge of the circle, surface states become less and less localized, till they merge with the extended states of the bulk and the uncertainty in position becomes infinite. 

Having said that, we can proceed to study the velocity renormalization, which occurs due to the squared momenta in $\Gamma_1$. That is, if we consider low momenta, then there is no velocity renormalization to first order, similarly to what has been observed in Refs.~\onlinecite{Diaz-Fernandez2017TuningField,Diaz-Fernandez2017Quantum-confinedJunctions}. Let us focus on the dispersion relation in the $k_x=0$ plane, where the dispersion is linear in the absence of electric field. The first order PT gives a term that typically increases the velocity if \eqref{eq:ModMin:CondS} is fulfilled (being the only range in which the expression is valid as we will discuss in the following subsections). The bands of the surface states are displaced, together with the bulk bands, due to the electric field. We shall denote the new position of the zero energy surface states as $k_{\mathrm{shift}}$ along the $k_y$ direction. With this definition, the dispersion relation within the first order PT is now
\begin{equation}  \label{eq:fMM:DispCones}
E_s (k_x=0) =  \eta \gamma_{0} 
+ \eta \zeta v_{f}^{\text{1PT}}\, \widetilde{k}_y + \mathcal{O} (f^2, k_y^2) \ ,
\end{equation}
where $\widetilde{k }_y \equiv k_y- k_{\mathrm{shift}}$. In addition $\gamma_{0}$ is a constant factor and the renormalized velocity reads
\begin{equation} \label{eq:fMM:vel1PT}
v_f^{\text{1PT}} \equiv v +  e f \left[\zeta  \frac{\partial \Gamma_{1}(0,k_y)}{\partial k_y} \right] _{k_{\mathrm{shift}}}\ .
\end{equation}
Notice that the velocity renormalization, up to first order PT, does not depend neither on $\zeta$ nor on $\eta$ because $k_{\mathrm{shift}}$ depends explicitly on chirality as $k_{\mathrm{shift}} = \zeta \abs{k_{\mathrm{shift}}}$. 

Next, we obtain the second order PT correction in a slab extended from $-w \leq z \leq w$ but infinite in the $X$ and $Y$ directions. Furthermore, we assume that the width $w$ is large enough to use the surface states obtained in the semi-infinite approximation. That is, the width is larger than the decay length of the surface states into the bulk so that states of opposite surfaces cannot hybridize. In a finite slab, the bulk states have the dispersion relation~\eqref{eq:ModMin:bEns} with a quantized momentum in the $Z$ direction due to the finite size of the system. The general form of the bulk states of the unperturbed Hamiltonian is
\begin{align}
    \psi_b^0 = \sqrt{\frac{1}{L_x L_y w}} e^{i q_x x} e^{i q_y y} \sin \left[ {q_z (z+w)}\right]  \Phi_b 
    \ , \label{eq:fMM:Bulk} 
\end{align} 
where $q_z={n \pi}/({2 w})$ with $n \in \mathbb{Z}$ and $\Phi_b$ is a constant and normalised spinor. The second order PT is given by
\begin{equation} \label{eq:fTM:Pert2}
    \delta E_s^{2} = \sum_{b} \frac{|\bra{\psi_s^0}H_f \ket{\psi_b^0}|^2}{E_s^0- E_b^0} \ ,
\end{equation}
where the sum index runs over all bulk states and $E_b^0$ ($E_s^0$) is defined according to Eq.~\ref{eq:ModMin:bEns} (Eq.~\ref{eq:ModMin:SurfE}). By implementing  \eqref{eq:fTM:Pert2} the second order correction obtained is
\begin{equation}
    \delta E_s^{2}  = - e^2 f^2 \eta \zeta \Gamma_2(k_x,k_y)\ ,
\end{equation}
with
\begin{equation}
\Gamma_2(k_x, k_y) \equiv \sum_{b} \phantom{}^{'} \frac{|I|^2}{w} 
\frac{2 v k_y}{(E_b^0)^2- (E_s^0)^2}\ ,
\end{equation}
where the primed sum runs over positive-energy bulk states due to the symmetry of the energy spectrum and  
$$
I\!=\!A_s\int_{-w}^{w}\!\! z \sin\left[q_z (z+w)\right] \left( e^{-\la_1 (\eta z + w)}-  e^{-\la_2 (\eta z + w)}\right)\,dz\ ,$$ 
only depends on the system parameters and $q_z$, but not on $\eta$. Notice that $\abs{E_s^0} < \abs{E_b^0} $ and therefore the second order correction $\delta E_s^{2}$ is always negative. Hence, the corrected energy dispersion up second order in the plane $k_x=0$, is given by
\begin{align} \label{eq:fMM:Epert2}
    E_s (k_x=0) &= \eta \zeta \left[v k_y - e^2 f^2 \Gamma_{2}(0,k_y)\right] \nonumber \\
      & + \eta e f \left[\Gamma_0 + \Gamma_{1}(0,k_y)\right] + \mathcal{O}(f^3) \ .
\end{align}

The contribution of the second order correction does not 
introduce new relevant effects except for changing $k_{\mathrm{shift}}$ and a introducing a reduction of the velocity as shown in the following expression
\begin{equation}
v_f^{\text{2PT}} \equiv  v +  \left[\zeta e f \frac{\partial \Gamma_{1}(0,k_y)}{\partial k_y} - e^2 f^2 \frac{\partial \Gamma_{2}(0,k_y)}{\partial k_y}
\right] _{k_{\mathrm{shift}}}\ .
\end{equation}

In summary, we identify two main effects of the electric field in the surface states: the shifting of the momenta of the surface states and the renormalization of the velocity. Since the type of surface state (A or B) depends on these two parameters, we expect that the electric field may induce a transition between different types of states. In the following, we discuss the main features that arise in the system under these conditions by comparing analytic treatment with the numerical calculations.

The shifting of the momenta of the surface states is the most salient result of the application of an external electric field. We refer to this effect as the \textit{shifting of the cone} of the dispersion relation of the surface states. Notice that the cone we are referring to is not a Dirac cone. Rather, it is the cone formed by each surface contributing states with opposite velocities in the field-free case. In a slab system, in the already mentioned limit of two decoupled surfaces, the accordance between the simulation results and the analytic ones is expected to be good as long as PT is valid. In a gapless system, an energy scale for PT is not so clear as in a gapped case. Moreover, the analytic expression of the surface states \eqref{eq:ModMin:SurfS} is valid only for momenta that fulfil Eq.~\eqref{eq:ModMin:CondS}. Hence, we expect PT to fail for electric fields that shift the momenta to values of ${k_x^2+ k_y^2} \simeq R^2$. In the following we refer to these momenta as the \textit{critical momenta}. PT results reflect intrinsically this range of validity. In fact, $\Gamma_1$ and $\Gamma_2$ have respectively a first and a second order pole at the critical momenta. From this perspective, it is clear that second order PT is going to fail faster than first order and the shift given by the constant $\Gamma_0$ will have a wider convergence radius.

\par Despite the aforementioned limitations, an expansion parameter for the PT can be defined. If we restrict the momenta of the surface states to $k_x=0$ and $k_y \ll R$, we can consider as the PT parameter, the ratio between the electric potential at the surfaces, $-ef \eta w $, and the gap between the surface and the bulk states at the critical momenta, $v\sqrt{m_0/m_1}$. In fact, it quantifies an effective gap compared with the displacement of the bands due to the electric potential. Hence, PT will be valid if $\abs{efw/(vR)} \ll 1$. For the parameters in Fig.~\ref{fig:fNa3Bi:PTvsSIM}, we found  that $f \ll \SI{12~(47)}{\milli\volt \nano \meter^{-1}}$ for $v= \SI{1~(4)}{\electronvolt \nano \meter}$, respectively. In the previously mentioned figure, we have plotted the 1PT up to values of 0.5 of the expansion parameter, while 2PT is plotted up to values of 0.1 finding a good accordance with the simulation results. 

\begin{figure}[htb] 
    \centering
    \includegraphics[width=0.50\textwidth]{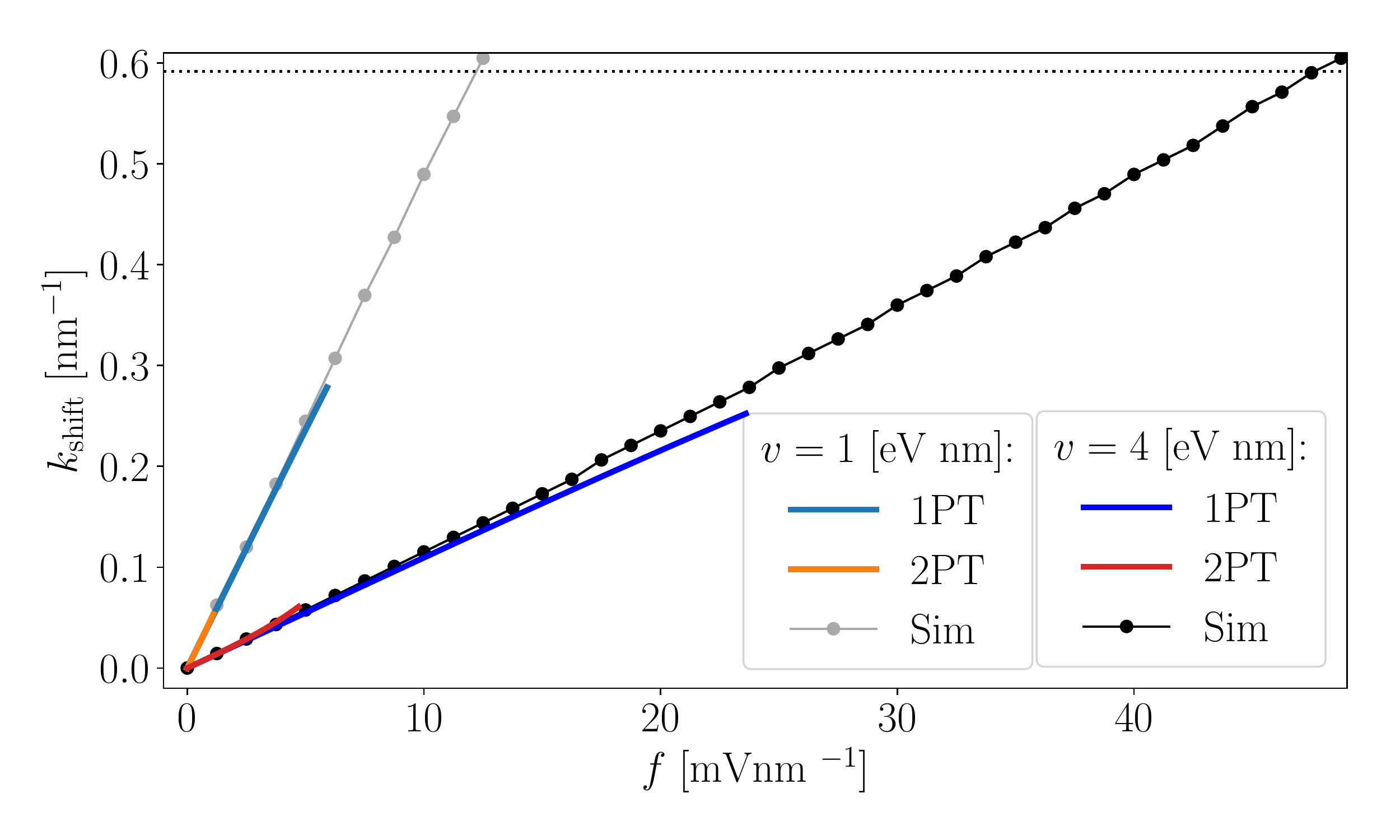}
    \caption{$k_{\mathrm{shift}}$ as a function of the electric field $f$ for a type A (B) slab of parameters $w=50 \si{~nm}$, $m_0=0.35 \si{~eV}$, $m_1=1.0 \si{~eV nm^2}$, $v= 1 \si{~eV nm}~ (4  \si{~eV nm} )$  and $\zeta=+1$; the dotted line is the critical momenta defined by $k_y = R$. 1PT denotes the first order PT, 2PT the results up to second order and Sim the simulations. We only calculate the PT up to relevant values of the small parameter of PT (see the main text for further details).}
    \label{fig:fMM:PTSIMkshift}
\end{figure}

As already mentioned, in order to quantify the shifting we define the shifted momenta $k_{\mathrm{shift}}$ as the momenta at which the branches intersect in the plane of $k_x=0$. It is obtained by finding the intersection of the two energy branches of the surface state dispersion  \eqref{eq:fMM:Epert2} at a fixed chirality. Figure \ref{fig:fMM:PTSIMkshift} shows a comparison of the PT result with the simulations: the second order starts to fail for really small electric fields and does not introduce relevant corrections, whereas the first order PT reproduces very well the simulation results. Therefore we will neglect the second order corrections in the following.  
\begin{figure}[htb]
    \centering
    \includegraphics[width=0.45\textwidth]{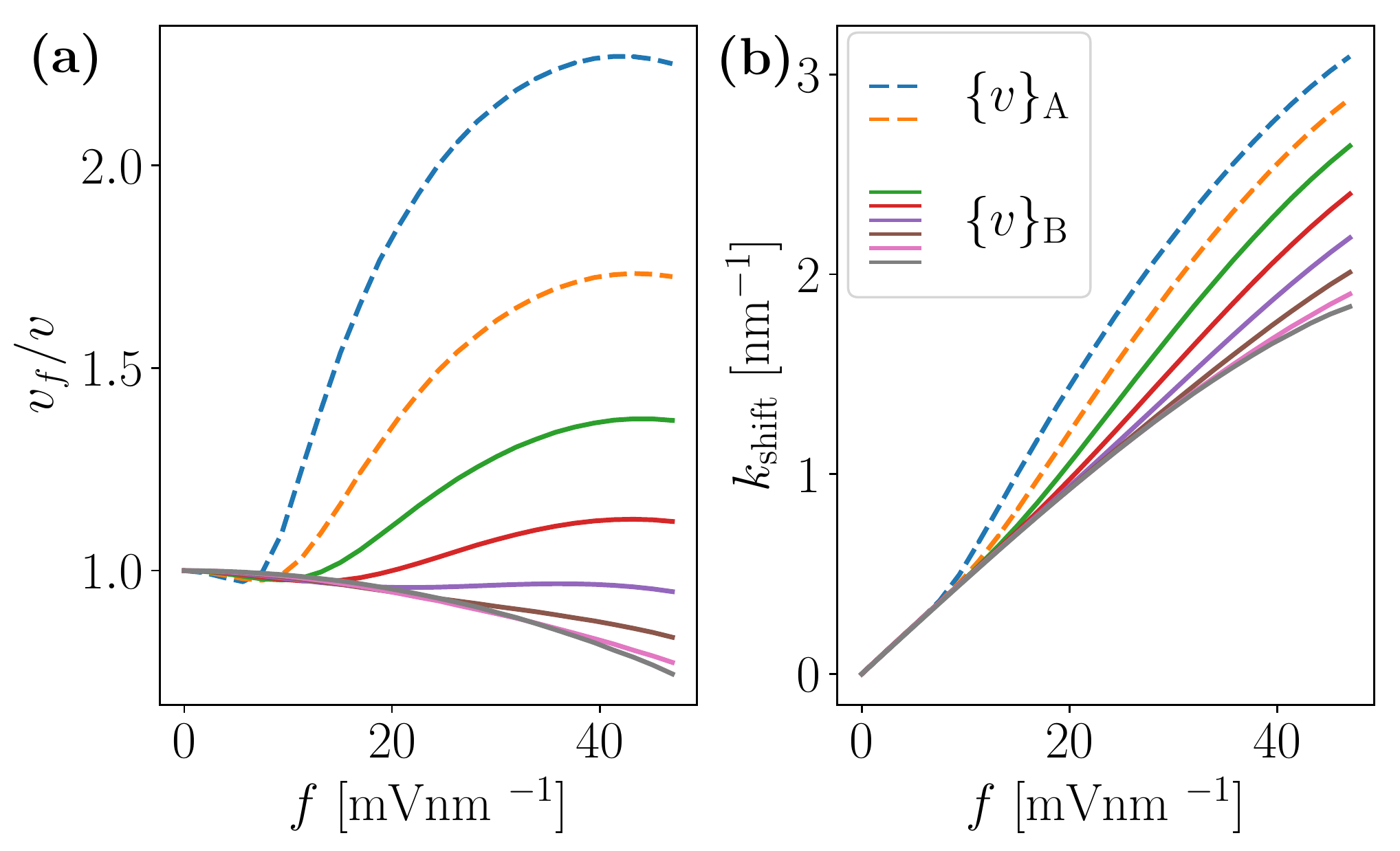}
    \caption{\textbf{(a)} Evolution of $v_f$ and \textbf{(b)} $k_{\mathrm{shift}}$ as a function of the electric field for different values of the parameter $v$ of the Hamiltonian \eqref{eq:ModMin:Ham}. The other parameters of the system are $w=50 \si{~nm}$, $m_0=0.35 \si{~eV}$, $m_1=1.0 \si{~eV nm^2}$ and $\zeta=+1$. The values of the velocities plotted are $\{v \}_{\mathrm{A}}= \{0.8, 1.0 \} \si{~eV nm}$ and $\{v \}_{\mathrm{B}}= \{1.2, 1.4, 1.6, 1.8, 2.0, 2.4 \} \si{~eV nm}$, meaning that for these values of $v$ the WSM is type A and B respectively.}
    \label{fig:fMM:VRen}
\end{figure}

Another aspect predicted by PT is the velocity renormalization. Since $v_f^{\mathrm{1PT}}$ is obtained as a derivative of $\Gamma_1$, it has a second order pole for the critical momenta (see Eq.~\eqref{eq:fMM:vel1PT}). Thus, the first order PT for the velocity converges within the same radius as the second order PT for $k_{\mathrm{shift}}$. In the minimal model, the radius of convergence of second order pole functions is utterly restricted to small $f$ (see Fig. \ref{fig:fMM:PTSIMkshift}), where a significant velocity renormalization is absent. Therefore, the effect can be only numerically studied in the non-perturbative regime. Figure \ref{fig:fMM:VRen} shows the evolution of the velocity and the $k_{\mathrm{shift}}$ with the electric field. Even if the PT is not valid for the regime of electric fields studied, it is worth mentioning that it captures important features of the effect as the independence of the results from the type of surface states.

\begin{figure}
    \centering
    \includegraphics[width=0.50\textwidth]{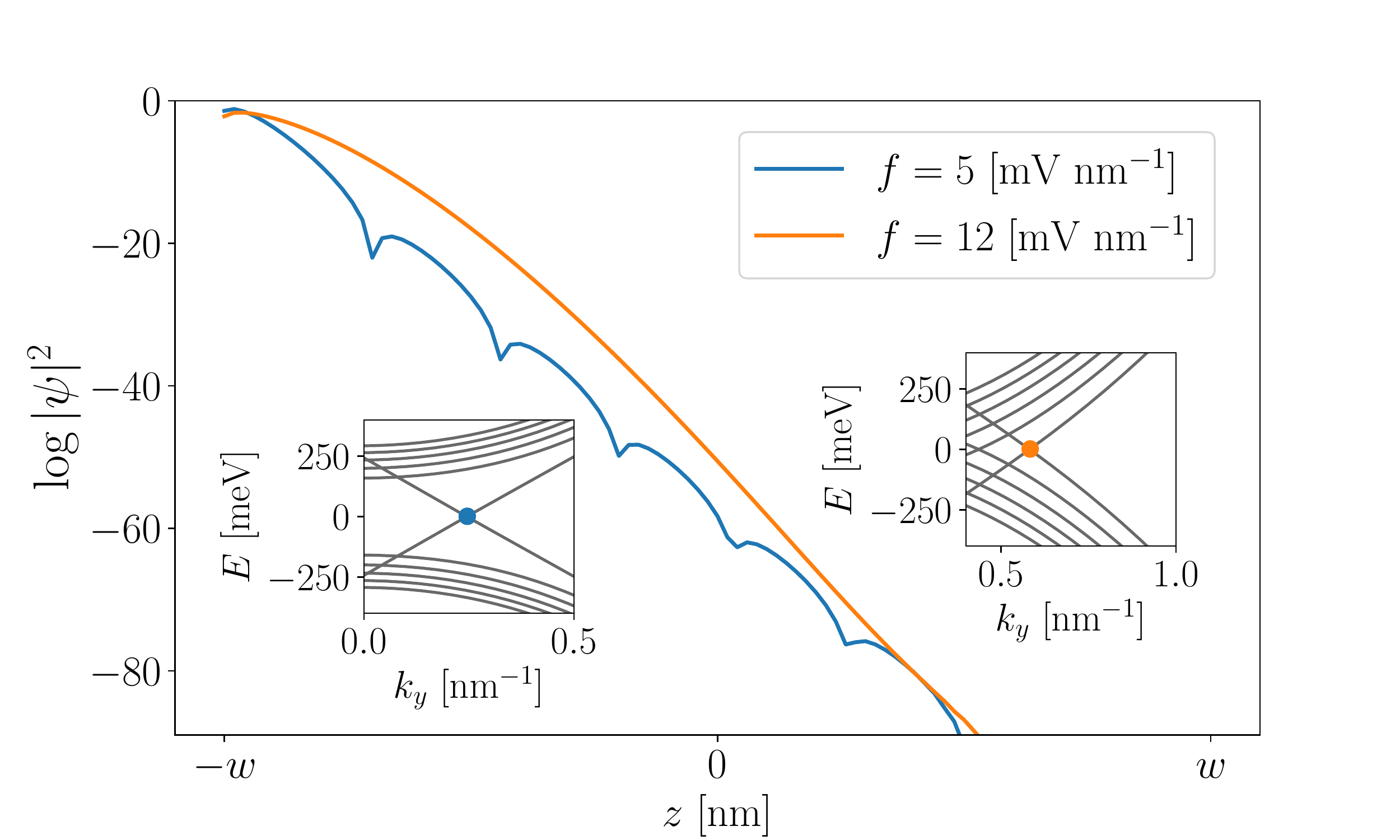}
    \caption{Type transition between type A and type B surface states for a system with parameters $w=50 \si{~nm}$, $m_0=0.35 \si{~eV}$, $m_1=1.0 \si{~eV nm^2}$, $v= 1 \si{~eV nm}$ and $\zeta=+1$. The main panel shows the wavefunctions and the insets show the first ten bands of the dispersion relation in the plane $k_x=0$. The left (right) inset corresponds to $f= 5 ~ (12) ~\si{mV nm^{-1}} $, the colored dots mark the energy and momentum of the wavefunctions plotted. Notice that the cone is shifted but the oscillatory decay is preserved for the small field whereas for the higher field the decay becomes purely exponential showing the type transition.}
   \label{fig:fMM:TransAtoB}
\end{figure}

The renormalization of the velocity and the shifting of the cone vertex leads to the possibility of a phase transition between types of surface states. In fact, the change in the velocity is directly related to the type transition because it modifies $r^2_{\mathrm{trans}}$ [see Eqs. \eqref{eq:ModMin:F} and \eqref{eq:ModMin:rtrans}]. Moreover, the shifting of the momenta of the surface states modifies the possible momenta that can be spanned in the left-hand side of \eqref{eq:ModMin:CondS}, typically increasing the terms on that side of the equation. To study the type transition, an accurate comparison of the evolution of $k_{\mathrm{shift}}^2$ and $r^2_{\mathrm{trans}}$ as a function of $f$ is needed. From the results exposed previously, the transition from type A to type B is expected: the shifting of the momenta is the most appreciable effect of the electric field, more noticeable than the velocity renormalization. Figure \ref{fig:fMM:TransAtoB} shows the transition $A \to B$ and the conversion of the oscillatory decay into a purely exponential one. 

\begin{figure}[tb]
    \centering
    \includegraphics[width=0.48\textwidth]{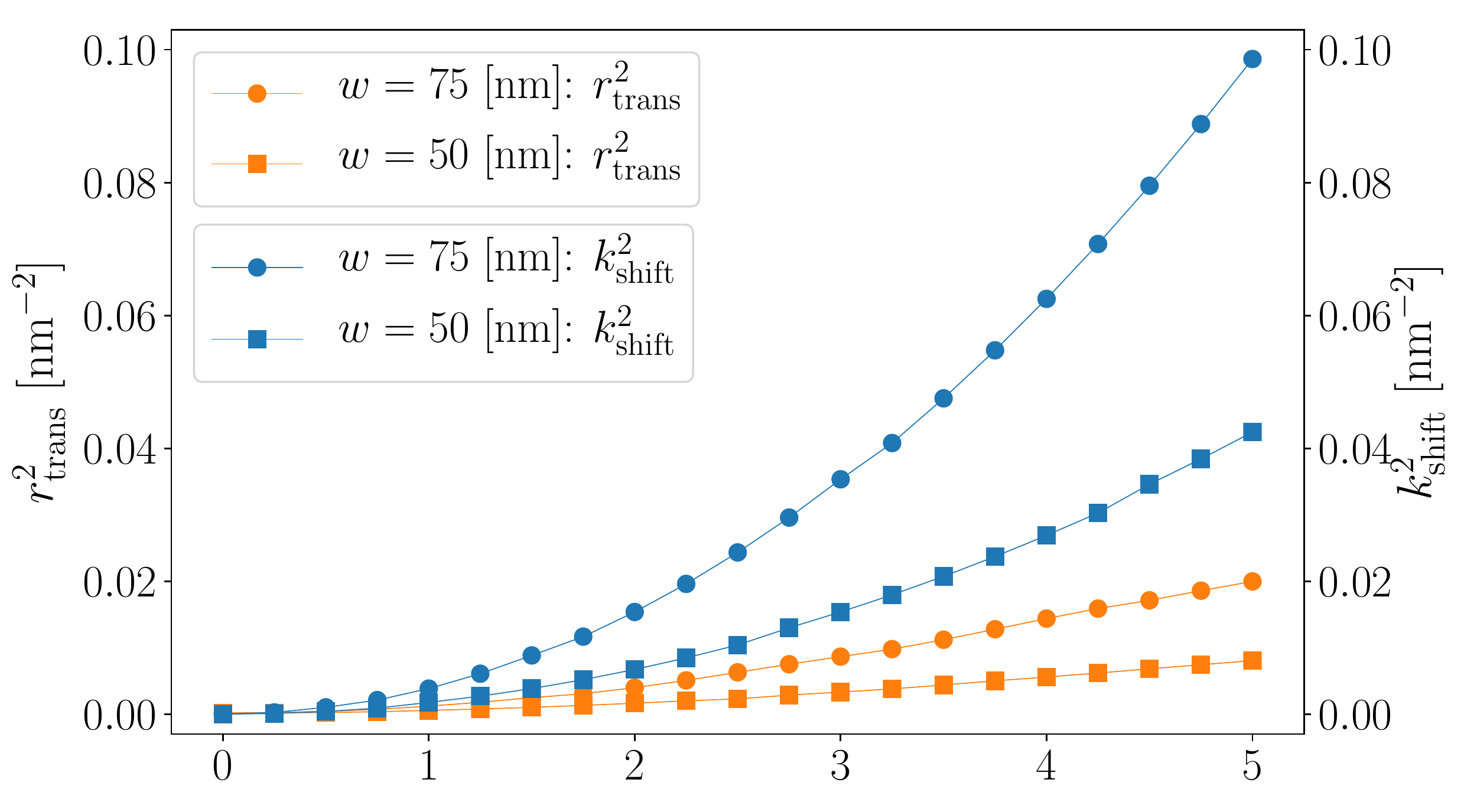}
    \caption{$r_{\mathrm{trans}}^2$ and $k_{\mathrm{shift}}^2$ as a function of the external field $f$; it is clear that the shifting in $k_y$ overtakes the condition that make possible the transitions $A \to B$. The parameters are $w= \{50, 75 \} \si{~nm}$, $m_0=0.35 \si{~eV}$, $m_1=1.0 \si{~eV nm^2}$, $v= v_{\mathrm{lim}} \simeq 1.18 \si{~eV nm} $.} \label{fig:fMM:Transition}
\end{figure}

The simulations confirm that the transitions $B \to A$ do not take place. To prove it, we simulate a type B WSM with $v= v_{\mathrm{lim}}$, where $v_{\mathrm{lim}} \equiv 2 \sqrt{m_0 m_1}$, corresponding to $r^2_{\mathrm{trans}}=0$. In this regime, a decreasing of the velocity turns the $r^2_{\mathrm{trans}}<0$ and allows for a type transition as long as $k_{\mathrm{shift}}^2$ does not increase beyond $r_{\mathrm{trans}}^2$.  In Fig. \ref{fig:fMM:Transition} the evolution of $k_{\mathrm{shift}}^2$ and $r_{\mathrm{trans}}^2$ are compared for two widths: the effect increases with $w$, as expected from the PT results, but the general behavior does not change with the thickness. 

\begin{figure}[htb]
    \centering
    \includegraphics[width=0.48\textwidth]{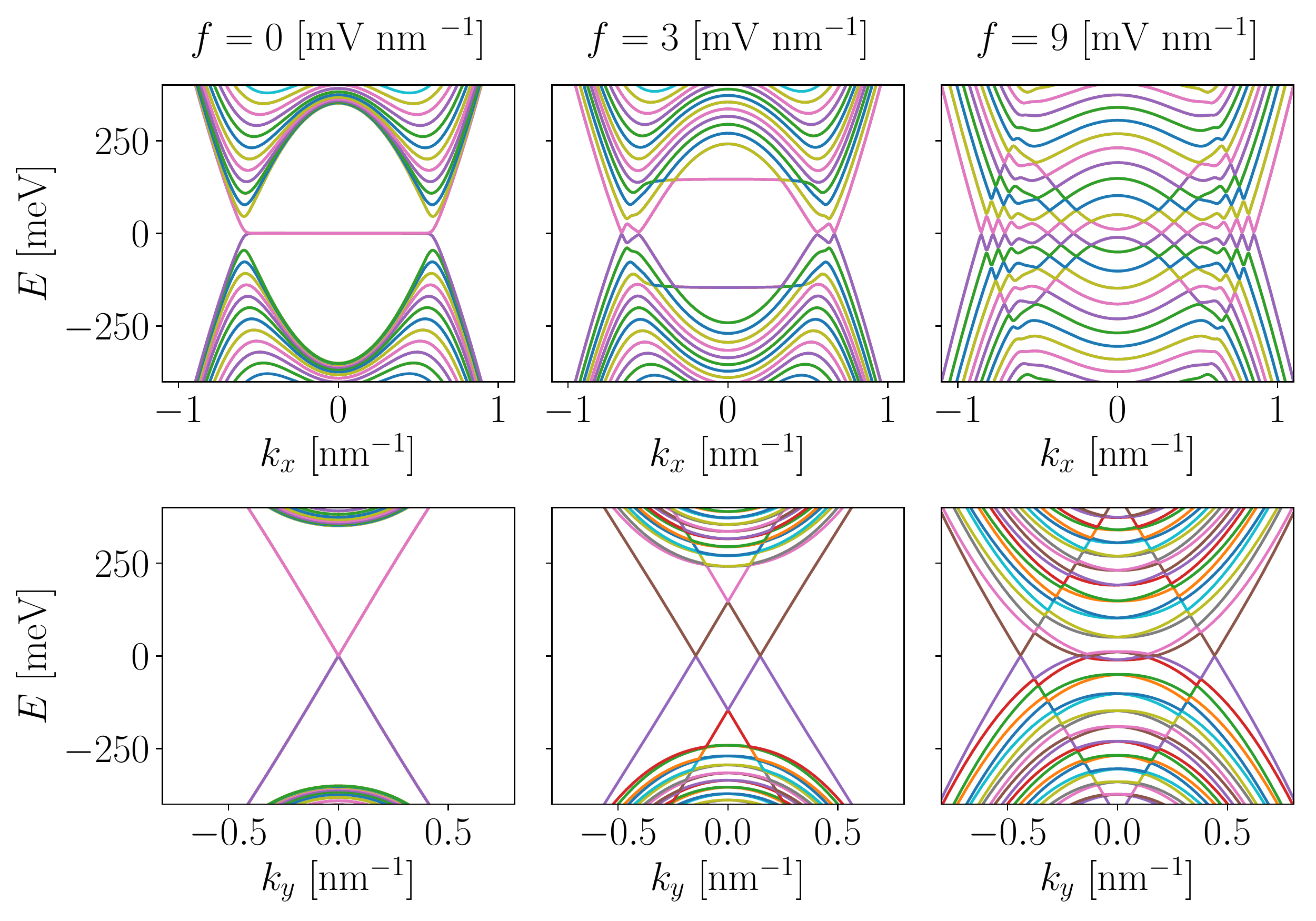}
    \caption{Dispersion relation of a DSM in the plane $k_y=0$ (upper) and $k_x=0$ (bottom) for a slab of $w=50 \si{~nm}$ with the same parameters of Fig. \ref{fig:ModMin:Disp}.}
    \label{fig:fMM_4x4}
\end{figure}

To complete our analysis of the minimal models, we study the case of the DSM with Hamiltonian \eqref{eq:HamD} comprising two copies of the minimal Weyl Hamiltonian \eqref{eq:ModMin:Ham} with different chiralities in each copy. In the absence of electric field, each boundary hosts two surface states with linear dispersion and opposite chirality. The electric field does not mix chiralities, but it breaks the spatial inversion symmetry leading to a splitting of the two cones of the opposite surfaces. In fact, the sign of the momentum shift induced by the electric field on the dispersion of the Fermi arcs depends on the chirality and $\eta$ (see Fig. \ref{fig:fMM_4x4}). The cited figure shows also the bands as a function of $k_x$ in the plane $k_y=0$: the effect of the electric field in the dispersion is the coalescence and distortion of the bands. This result is expected as the crossing of the cones is shifted to non-zero $k_y$. It is important to notice that these cones are not the Dirac cones, rather they correspond to the intersection of the two surface states of the same chirality and opposite surfaces near the zero energy, as mentioned earlier in the text. In fact, the actual Dirac cones actually shift up or down in energy, depending on the surface they are located at. 

\revision{Before concluding this section, it is interesting to notice that, as it can be seen in Fig.~\ref{fig:fMM_4x4}, the surface states linear branches directly enter into the bulk bands and remain unaltered despite the presence of the latter. This means that surface states and bulk states do not hybridize. We have observed this fact also by looking at the evolution of the surface states as the field is increased. To this respect, more details can be found in the Supplemental Material~\cite{Suppl}.
It is important to stress, however, that the bulk states are modified by the electric field as well. States of the conduction (valence) band transition from being completely extended through the system to localize towards regions of lower (higher) electric potential. This is in fact the reason why bulk and surface states do not hybridize, because the former localize at regions where the latter have negligible probability density.}

\subsection{Model for the Dirac semimetal \ce{Na_3Bi}}

Throughout this section, we extend the previous results of the minimal model to the Hamiltonian for \ce{Na_3Bi}. Since the electric field does not mix the chiralities, we study the $2\times 2$ model with nondegenerate Weyl nodes given by Eq.~\eqref{eq:Na3Bi:HamW}. We follow the same approach as with the minimal model up to first order PT applying Eq.~\eqref{eq:fMM:Pert1st} to the surface states defined by \eqref{eq:Na3Bi:SurfS}. In order to follow the shifting of the cones we set $k_z=0$ and to simplify the notation, we define all the quantities at zero $k_z$. The first order PT correction now is
\begin{equation} \label{eq:fNa3Bi:Epert1}
    \delta E^1_s = e f \eta \left[ \Gamma_0 + \Gamma_1 (k_x) \right]\ ,
\end{equation}
with $\Gamma_0$ a constant obtained from the model parameters and $\Gamma_1(k_x)$ a function of $k_x$
\begin{subequations}
\begin{align} \label{eq:fNa3Bi:Gamma0}
\Gamma_0 & =  - w +\frac{1}{2 \Delta} + \frac{\Delta}{R^2 - k_{x,0}^2}  \ ,\\    \label{eq:fNa3Bi:Gamma1}
\Gamma_1(k_x) &= \frac{\Delta  k_x (k_x+2   \eta \zeta k_{x,0})}{\left(R^2- k_{x,0}^2\right) \left[R^2-(k_x+ \eta \zeta  k_{x,0})^2\right]}\ .
\end{align}
\end{subequations}

Before proceeding further, a couple of words must be said about the convergence of PT in this case. As for the minimal model, the surface states \eqref{eq:Na3Bi:SurfS} exist if $F< \Delta^2$ \cite{Benito-Matias2019}. The former condition implies now that 
\begin{equation} \label{fNa3Bi:kc}
    \big(k_x+\zeta  \eta  k_{x,0}\big)^2+ \frac{m_1}{m_2}k_z^2<R^2\ .
\end{equation}
These domains represent ellipses in the $k_x-k_z$ plane and imposing $k_z=0$ they are a constraint for the values of $k_x$ that must fulfil $k_{x,c}^-< {k_x}< k_{x, c}^+$, where $k_{x,c}^\pm$ is a function of the model parameters and is obtained from \eqref{fNa3Bi:kc}.
We expect PT to converge in the range $k_{x,c}^- \ll {k_x} \ll k_{x, c}^+$.

\begin{figure}[htb]
    \centering
\includegraphics[width =  0.46\textwidth]{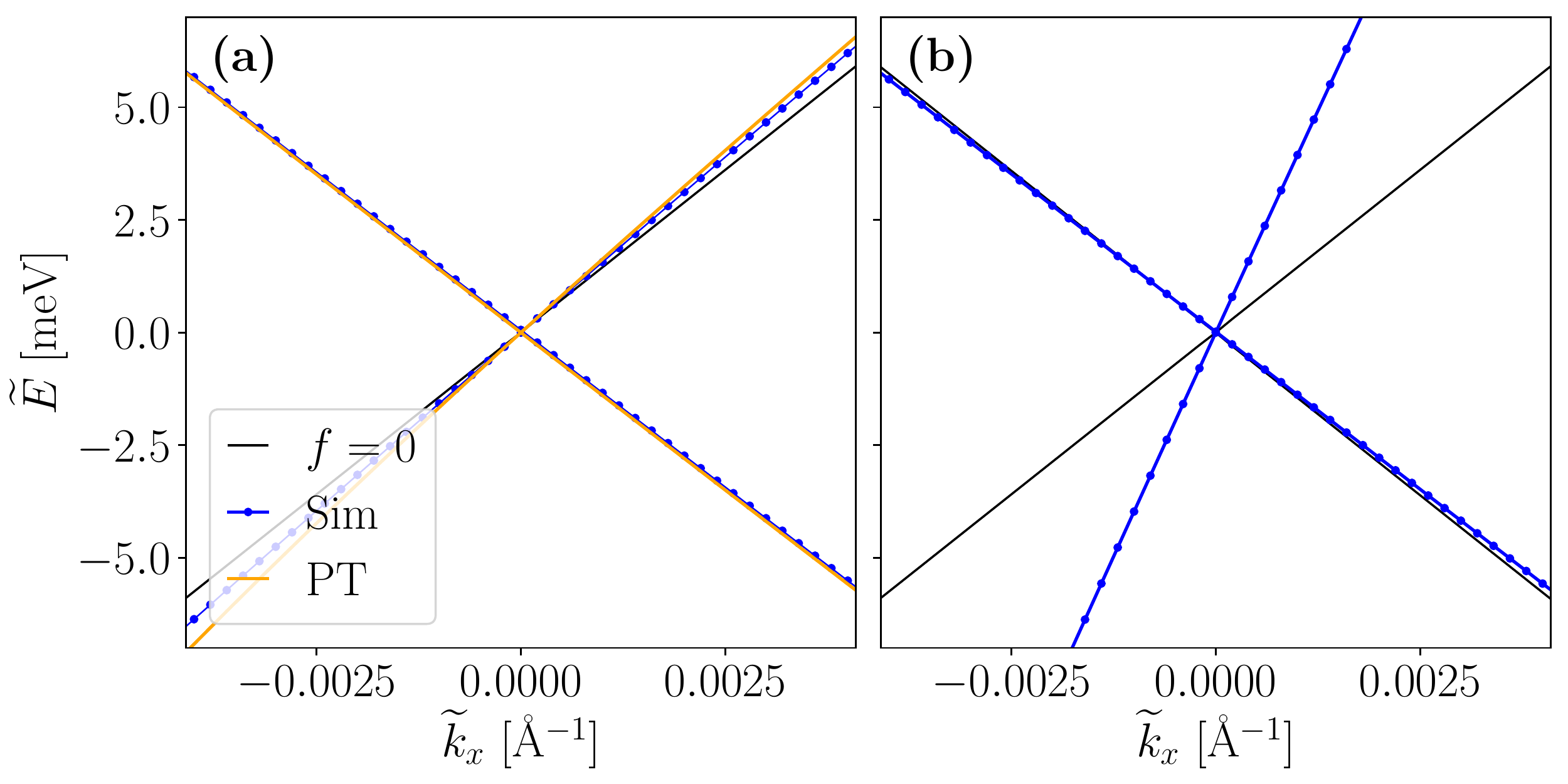}
    \caption{Reshaping of the cones due to the electric field: $f=0.08 \si{~mV \angstrom^{-1}}$ in \textbf{(a)} and $f=1.0 \si{~mV \angstrom^{-1}}$ in \textbf{(b)}. 
    The cones are centred in energy and momenta using the variables $\widetilde{k}_x \equiv k_x - k_{\mathrm{shift}}$ and $\widetilde{E} \equiv E - E_{\mathrm{shift}}$. PT results and simulations are compared for small $f$ finding a good accordance. The width of the system is $w = \SI{200}{\angstrom}$ and $\zeta = 1$, the cones in the absence of electric field are plotted in black.}
    \label{fig:fNa3Bi:Cones}
\end{figure}

In order to further simplify the problem and have an intuitive idea of the phenomena, it is interesting to study the behavior of the energy dispersion for small momenta by expanding the perturbed energy in $k_x \ll 1$ as follows
\begin{multline} \label{eq:Na3BiE1Series}
    E_s (k_z=0)
    =C_1+ e f \eta \big( {2 \Delta^{-1}} + \gamma_0 - w \big) \\
    +\zeta (\eta v C_3 + e f \gamma_1) k_x 
    + \eta e f \gamma_2 k_x^2
    + \mathcal{O}\left(f^2, k_x^3\right),
\end{multline}
where $\gamma_i ~(i=0,1,2)$ are positive constants. Moreover the renormalized velocity is defined as previously by way of an expansion near the shifted cones as follows:
\begin{equation} \label{eq:fNa3Bi:velRen}
    v^{\text{1PT}}_f \equiv v +  e f \left[\zeta  \frac{\partial \Gamma_{1}}{\partial k_x} \right] _{k_{\mathrm{shift}}}\ .
\end{equation}
From inspection of Eq.~\eqref{eq:Na3BiE1Series} we observe  some features revealed by the minimal model: the bands are displaced by a constant term and they are transformed by the terms $\gamma_{1}$ and $\gamma_2$. However, this scenario presents important modifications with respect to the minimal model.
First of all, $\Gamma_1$ depends on the chirality $\zeta$ and $\eta$, therefore the renormalized velocity is asymmetric. Using the numerical values of table \ref{tab:Na3Bi} and the PT expression for $v^{\text{1PT}}_f$ \eqref{eq:fNa3Bi:velRen}, we find that the velocity increases for surface states with $\eta \zeta = + 1$ and decreases for $\eta \zeta = -1$. This phenomenon is perfectly observed also in the simulations, as seen in Fig. \ref{fig:fNa3Bi:Cones}. 

\begin{figure}
    \centering
    \includegraphics[width =  0.5\textwidth]{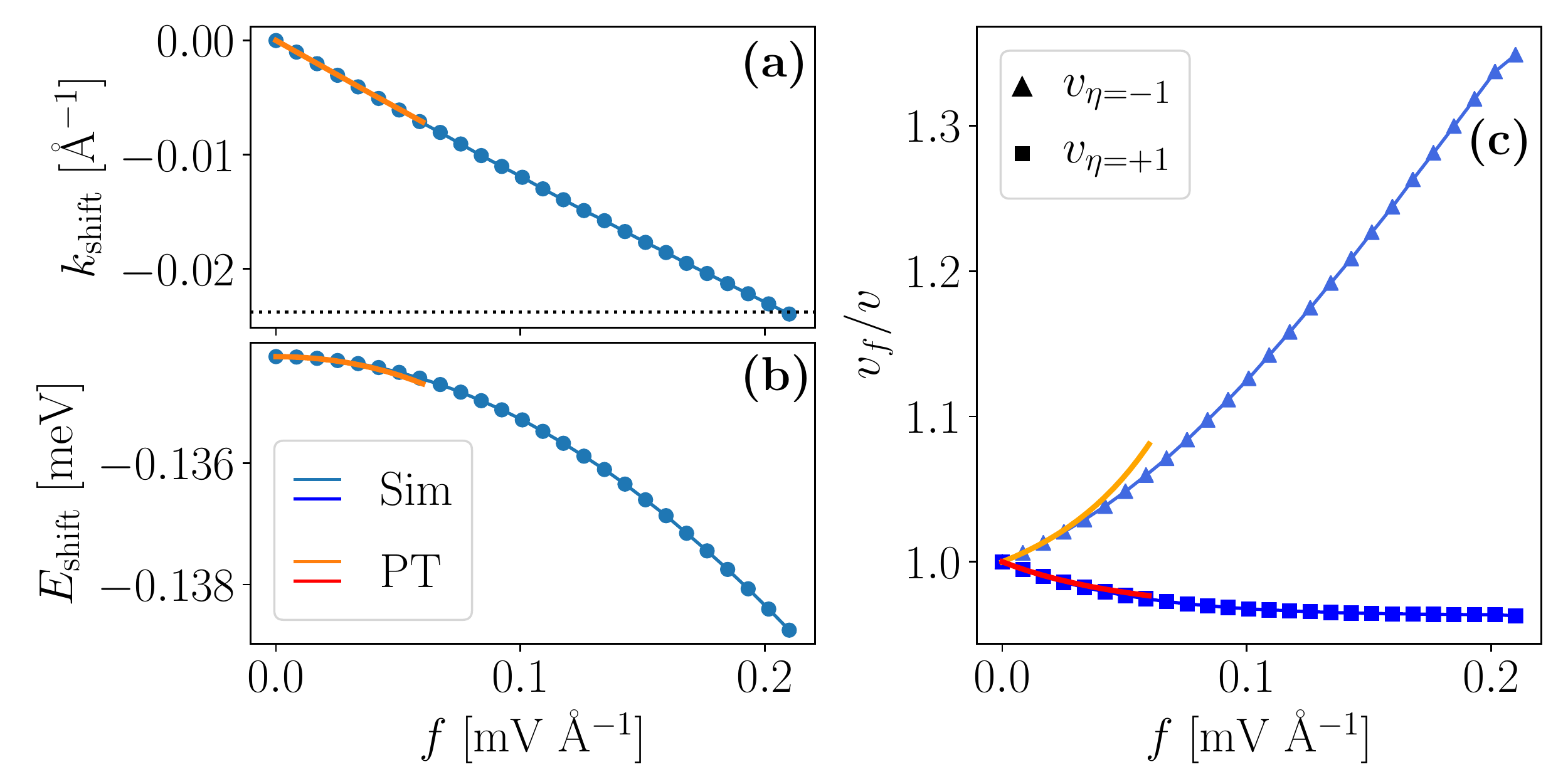}
    \caption{Comparison between PT results and numerical simulation for a \ce{Na_3Bi} slab of $w= 200~ \si{\angstrom}$. \textbf{(a)} shows $k_{\mathrm{shift}}$, \textbf{(b)} the $E_{\mathrm{shift}}$ and  \textbf{(c)} the velocity as a function of the electric field. The dashed line marks the minimal critical momenta $k_c$. }
    \label{fig:fNa3Bi:PTvsSIM}
\end{figure}

Moreover, the bands of the surface states are not flat and they do not cross at zero energy for $k_z=0$. Therefore, the shifting produced by the electric field modifies the crossing of the two branches in momenta and in energy what defines a new $k_{\mathrm{shift}}$ and $E_{\mathrm{shift}}$. The evolution of these two quantities is plotted in Fig. \ref{fig:fNa3Bi:PTvsSIM} along with the velocity renormalization. The previously mentioned figure compares PT with the numerical simulations finding a good accordance as long as the shifted momenta fulfill the already discussed restrictions. We find that the PT converges if $f \lesssim 10^{-2}~\si{\milli \volt \angstrom^{-1}}$ for systems with $w \sim 10^2~\si{\angstrom}$. Remarkably, despite considering weak electric fields, the effect on the velocity renormalization is not negligible at all.

\begin{figure}
    \centering
    \includegraphics[width =  0.49\textwidth]{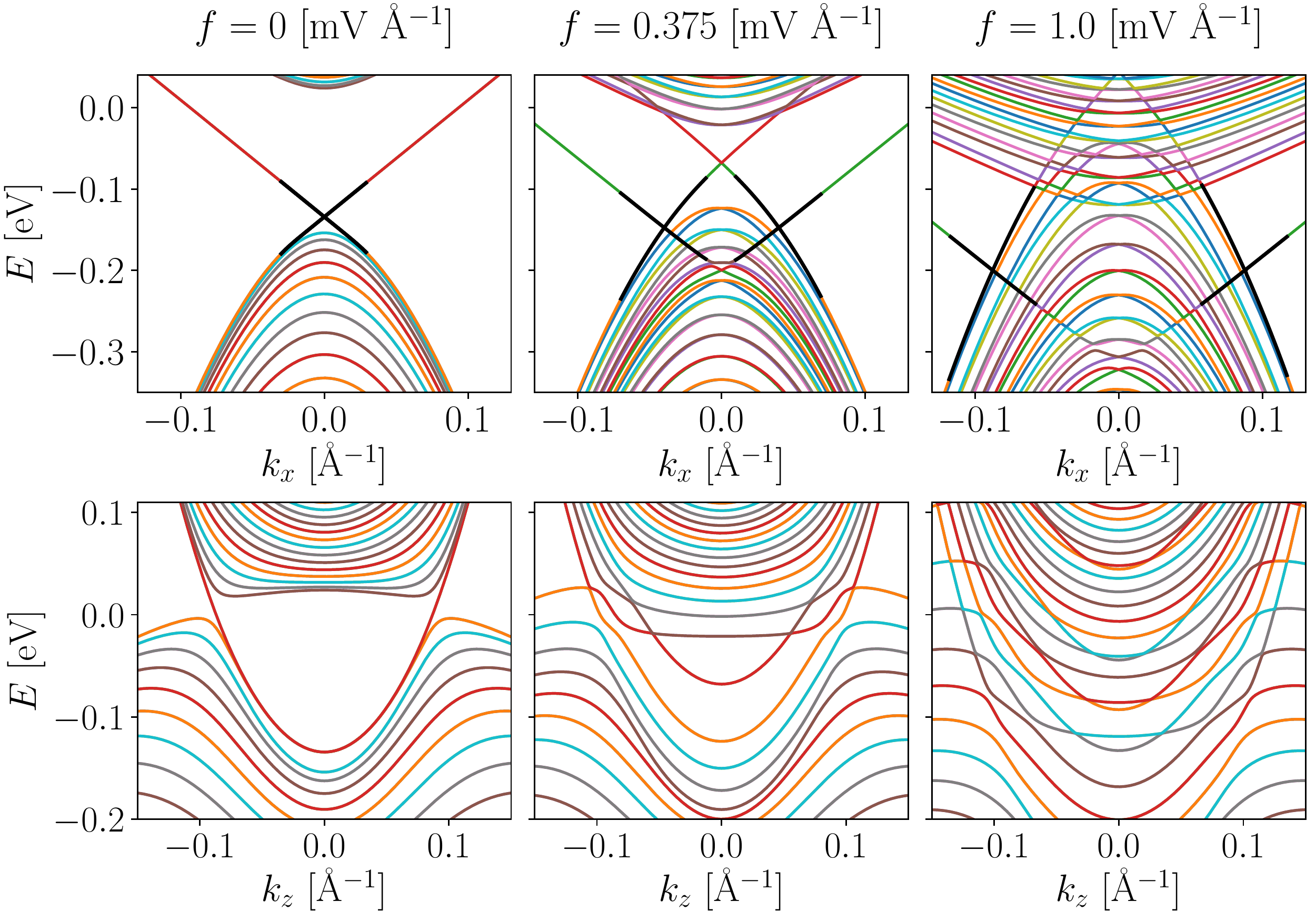}
    \caption{Change of the dispersion relation with the electric field $f$ in a slab of $w= \SI{200}{\angstrom}$ and setting $\zeta = +1$. In the upper panels, the shifting of the cones is utterly visible in the dispersion as a function of $k_x$ (at $k_z=0$). In the bottom panels, it can be seen the coalescence of the bands in the dispersion as a function of $k_z$ (at $k_x=0$). To improve visibility, we have underlined the crossing of the surface energy branches with a black line.}
    \label{fig:fNa3Bi:Disp}
\end{figure}

Finally, let us pay attention to the dispersion relation for the complete Hamiltonian, i.e. the DSM Hamiltonian, as plotted in Fig. \ref{fig:fNa3Bi:Disp}. The effect is analogous to the one found in the minimal model, in that the branches  in black move away from each other in momentum as the bulk states approach in energy.

\section{Experimental proposal} \label{sec:proposal}

\begin{figure}[htb]
    \centering
    \includegraphics[width =  0.48\textwidth]{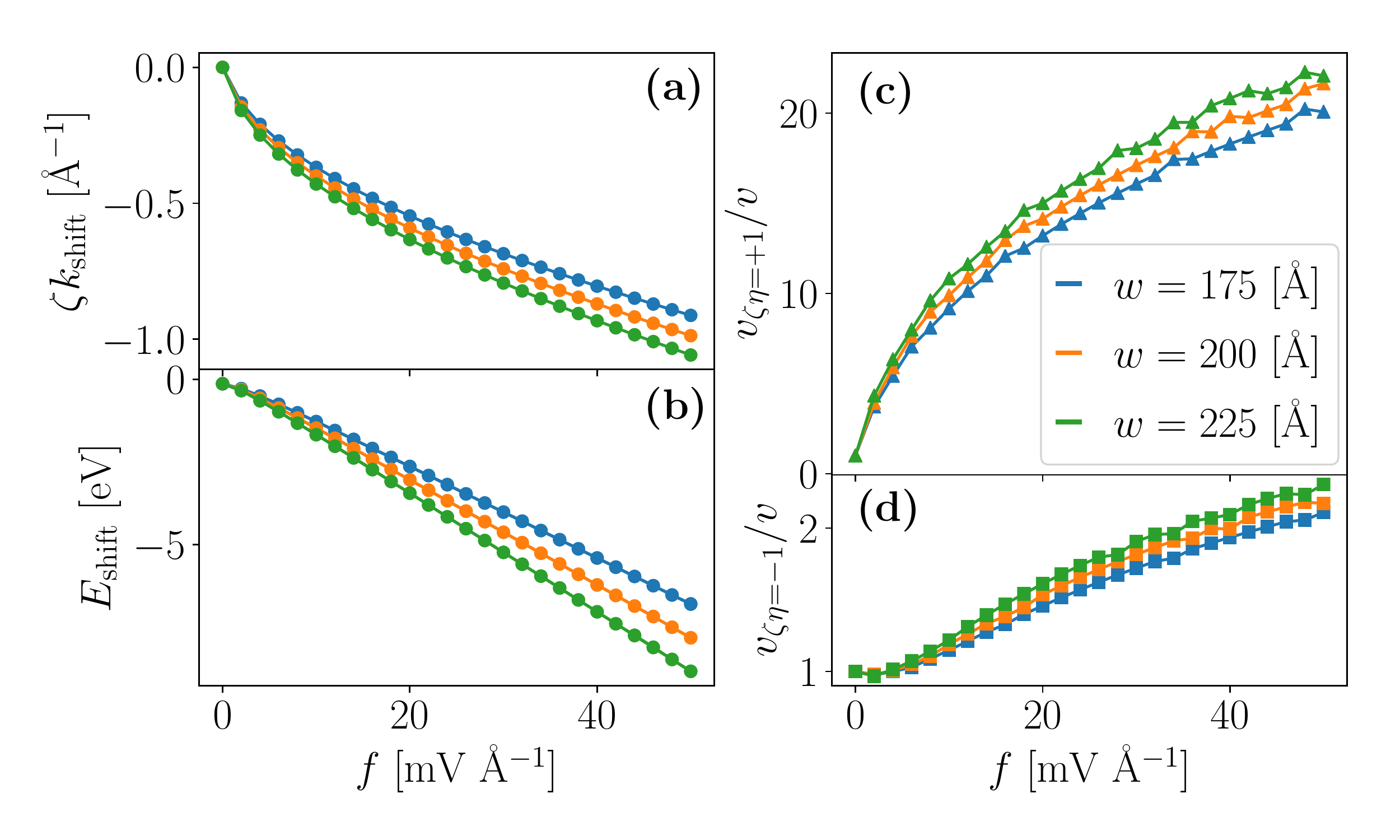}
    \caption{\textbf{(a)} $k_{\mathrm{shift}}$, \textbf{(b)} $E_{\mathrm{shift}}$ and \textbf{(c)}-\textbf{(d)} velocity as a function of the external electric field for three different widths of a \ce{Na_3Bi} slab. }
    \label{fig:fNa3Bi:Sim}
\end{figure}

A very recent paper has reported experimental measurements in the presence of an electric field in ultra thin \ce{Na_3Bi} films \cite{Collins2018}. The cited article, based on a theoretical proposal~\cite{Pan2015}, studies a different regime from the present work: the system comprises few-layer \ce{Na_3Bi(001)} films grown in the $Z$ direction, along which the electric field is applied. Its main result is the closing and opening of a gap such that topological and trivial phases are induced. The importance of this experimental realisation is relevant from the standpoint of applications of the present work. In fact, it assures the possibility of realization of the proposed scenario and gives a magnitude of the electric fields that can be implemented experimentally. In the ultrathin film set up, the electric fields were implemented using two methods: doping the surface with potassium and with scanning tunnelling spectroscopy varying the tip–sample separation. The obtained electric fields are of the order of a few $\si{V nm^{-1}}$ in a sample of atomic width.  In the system proposed in this work, the thickness of the sample must be of the order of $10^2 \si{\angstrom}$ in order to have two decoupled surfaces. In the literature these samples are labelled as thin films and have been grown by Molecular Beam Epitaxy for \ce{Na_3Bi} \cite{Hellerstedt2016, Zhang2014} and \ce{Cd_3As_2} \cite{Schumann2018}. Due to this estimated thickness, we expect that lower fields can be achieved. Indeed, even for electric fields one order of magnitude lower than the ones obtained by~ \citeauthor{Collins2018} \cite{Collins2018}, we find that the velocity renormalization is a non-negligible effect, as seen in Fig. \ref{fig:fNa3Bi:Sim}.

\section{Conclusions} \label{sec:conc}

In summary, we have obtained a suitable method to displace the Dirac nodes of the surface states in the BZ with a tunable external electric field. Not only the position of the cones is modified but also the Fermi velocity can be altered. Moreover, both effects depend on the chirality of the node at a given surface. The renormalization of the Fermi velocity would have a direct impact on the electronic transport properties through the surface of the semimetal thin films. In the case of \ce{Na_3Bi} we show that these effects would be quite significant and due to their chiral dependence, we envision the possibilities for applications to chiral electronic devices \cite{Kharzeev2013}. 

For some range of parameters, the renormalization of the Fermi velocity induced by the external electric field implies a transition from type A, with oscillatory decay and very short decay lengths, to type B surface states, with longer decay lengths and pure exponential decay into the bulk. This may lead to a very large increase in the hybridization of the two surfaces in a thin film. The coupling of the opposite surfaces is a necessary ingredient of the recently observed 3D Quantum Hall effect based on Weyl orbits~\cite{Zhang2019,Nishihaya2019} and we foresee important observable consequences of this transition in quantum Hall transport experiments with topological semimetals. 

\acknowledgments

We acknowledge financial support through Spanish grants PGC2018-094180-B-I00 (MCIU/AEI/FEDER, EU), FIS2015-63770-P and MAT2016-75955 (MINECO/FEDER, EU), CAM/FEDER  Project  No.S2018/TCS-4342 (QUITEMAD-CM) and CSIC Research Platform PTI-001.  

\bibliography{refs}

\end{document}